\documentclass[letterpaper, 10 pt, conference]{ieeeconf}  % Comment this line out if you need a4paper

\IEEEoverridecommandlockouts                              % This command is only needed if 
\usepackage{graphicx} % for pdf, bitmapped graphics files
\title{\LARGE \bf
Regional Transportation Modeling for Equitable Electric Vehicle Charging Infrastructure Design}

\author{Ismaeel Babur and Jane Macfarlane, Institute of Transportation Studies, UC Berkeley }% <-this % stops a space

\begin{document}

%\thanks{Both authors are with the Institute of Transportation Studies, UC Berkeley}% <-this % stops a space

\maketitle
\thispagestyle{empty}
\pagestyle{empty}

%%%%%%%%%%%%%%%%%%%%%%%%%%%%%%%%%%%%%%%%%%%%%%%%%%%%%%%%%%%%%%%%%%%%%%%%%%%%%%%%
\begin{abstract}

The widespread adoption of battery electric vehicles (BEVs) holds promise for mitigating emission-related health impacts, particularly for low-income communities disproportionately affected by exposure to traffic-related air pollution. However, designing effective charging infrastructure necessitates a regional modeling approach that accounts for the inherent cross-jurisdictional nature of mobility patterns. This study underscores the importance of regional modeling in optimizing charging station deployment and evaluating the environmental justice implications for equity priority communities. We present a large-scale regional transportation modeling analysis leveraging Mobiliti, a cloud-based platform that employs parallel discrete event simulation to enable rapid computation. Our approach identifies the spatial demand density for charging infrastructure by analyzing over 19 million trips in the San Francisco Bay Area and determining the threshold points where BEVs may require charging across a typical day. By transitioning these trips that originate outside equity priority communities to BEVs, we quantify the potential emission reductions within these vulnerable areas. The regional modeling framework captures the complex interactions between travel behavior, vehicle characteristics, and charging needs, while accounting for the interconnectivity of infrastructure across municipal boundaries. This study demonstrates the critical role of regional modeling in designing equitable BEV charging networks that address environmental justice concerns. The findings inform strategies for deploying charging infrastructure that maximizes accessibility, minimizes range anxiety, and prioritizes the health and well-being of communities disproportionately burdened by transportation emissions.

\end{abstract}

%%%%%%%%%%%%%%%%%%%%%%%%%%%%%%%%%%%%%%%%%%%%%%%%%%%%%%%%%%%%%%%%%%%%%%%%%%%%%%%%
\section{Introduction}

The deployment of infrastructure for charging battery electric vehicles (BEVs), beyond the typical work home model, entails several complex challenges. Spatial demand forecasting involves accurately predicting where charging infrastructure is needed, considering factors such as travel patterns, trip lengths, and energy consumption across different regions and demographics. Grid capacity and integration require significant upgrades to accommodate the increased electrical load from widespread BEV charging while ensuring the reliability and resiliency of the grid. Site selection involves balancing factors such as accessibility, land availability, zoning regulations, and community impacts to identify optimal charging station locations. Charging speed and standards necessitate a mix of charging station types to cater to different needs, along with standardization of charging protocols and connectors across vehicle manufacturers. Cost and financing challenges include substantial upfront capital costs for hardware, installation, and grid upgrades, which require sustainable financing models and incentives for widespread adoption. In addition, equity and accessibility concerns must address potential disparities in infrastructure deployment across socioeconomic and demographic groups to ensure equitable access and avoid exacerbating existing inequalities.
 
The design of efficient charging infrastructure for battery electric vehicles (BEVs) requires a regional modeling approach that accounts for the inherently broad geographic nature of travel patterns. Regional transportation modeling on a large scale can provide invaluable insights into the complex spatial dynamics of charging demand.
Individual trips routinely transcend local boundaries, spanning multiple jurisdictions and communities within a region. Trip lengths, origins, destinations, and route choices exhibit substantial variability that extends beyond municipal limits. Regional modeling enables capturing this diversity in travel behavior and accounting for inter-community mobility flows. In addition, the energy consumption and range requirements of BEVs depend on a multitude of factors, including vehicle specifications, driving conditions, and route characteristics (e.g., terrain, traffic congestion). Regional modeling facilitates the incorporation of these variables in an expansive geographic area, yielding a more comprehensive understanding of the spatial distribution of charging needs. Crucially, deploying an effective charging network requires a coordinated regional approach that considers the interconnectivity and accessibility of charging stations across a broad area. Regional modeling aids in identifying optimal site locations that enable seamless long-distance travel, minimize range anxiety, and maximize the overall utility of the infrastructure network.

Furthermore, integrating charging infrastructure with the electrical grid and ensuring adequate capacity necessitates close collaboration with utility providers and grid operators at a regional scale. Regional modeling informs grid capacity planning, infrastructure upgrades, and load management strategies across the broader service area. Implementing a robust, equitable charging network often necessitates coordination among diverse stakeholders, including local governments, transportation agencies, utility providers, community organizations, and regional planning bodies. Regional modeling facilitates this multi-stakeholder collaboration, aligning efforts and strategies across jurisdictional boundaries to create a cohesive regional vision.

By adopting a regional modeling approach, transportation planners, policymakers, and stakeholders can develop a comprehensive understanding of the spatial charging demand, optimize the deployment of charging infrastructure, and facilitate the widespread adoption of BEVs while addressing regional challenges and leveraging economies of scale.

To add to the complexity, adverse environmental and public health impacts associated with road transportation are becoming increasingly important research and policy priorities as urbanization continues in metropolitan areas. Reduction of traffic congestion, accidents, and multi-modal pollution (including noise, water, and air contamination) must be examined and mitigated through equitable, evidence-based approaches. These transportation-related impacts are experienced in many ways, ranging from the individual road user level to broader societal and environmental realms. Notably, traffic congestion and air pollution represent both highly visible and less directly observable consequences of road transportation, respectively.

Navigation applications that implement personalized routes designed to improve travel times may in fact exacerbate traffic congestion and environmental pollution, particularly on parts of the road network that were not intended to handle high density traffic. Moreover, as advancements in vehicle connectivity and automation enable increasingly tailored routing solutions geared towards individual speed and convenience, the implications of such optimization efforts must be scrutinized not only at the macroscopic system level but also within localized neighborhoods and at the individual level. 

Of particular importance is the consideration of localized air quality assessments in addressing environmental justice concerns, particularly in areas characterized by a concentration of high-capacity roadways where lower-income and otherwise disadvantaged populations tend to reside. Exposure to transportation-related air pollutants poses significant public health risks, including cardiovascular and respiratory diseases, cancer, and adverse birth outcomes, underscoring the importance of understanding and mitigating the impacts of transportation-related pollution on vulnerable communities.

This study uses a regional simulation model, Mobiliti \cite{chan_mobiliti_2018}, to investigate the equity implications of Battery Electric Vehicle (BEV) introduction and its potential to mitigate emissions in disadvantaged communities, which often bear disproportionate exposure to traffic-related air pollution. 

The vehicle routing assignments simulated in Mobiliti generate link-level estimates of vehicle flows, speeds, and fuel consumption across the San Francisco (SF) Bay Area road network, composed of over 1 million links. These high-resolution transportation metrics enable the estimation of localized congestion and pollution impacts, which are then evaluated in the context of equity priority communities.

The following sections present our modeling approach for characterizing energy use patterns of typical battery electric vehicles (BEVs) and leveraging a regional-scale transportation simulation platform to generate large-scale mobility datasets. We first introduce simplified energy consumption models for BEVs and compare them to a more complex vehicle dynamics model with finer granularity. Next, we showcase the capabilities of our high-performance Mobiliti simulation platform to generate large-scale trip datasets encompassing over 19 million trips across the San Francisco Bay Area region. Using this extensive dataset, we model the spatial demand density for charging infrastructure, assuming a hypothetical scenario where a reasonable percentage (based on known market share) of vehicles are transitioned to BEVs.
Building upon these regional mobility and energy demand insights, we then assess the potential positive impacts of widespread BEV adoption on air quality in equity priority communities residing adjacent to high-traffic areas. By transitioning pass-through traffic to zero-emission BEVs, we hope to quantify the localized reductions in transportation-related emissions that disproportionately burden these vulnerable neighborhoods.
Finally, we emphasize the critical importance of employing regional transportation modeling frameworks in the planning and deployment of BEV charging infrastructure. Beyond informing optimal site selections and network configurations, these robust modeling tools enable holistic evaluations of equity implications and environmental justice considerations.
Our analysis underscores the necessity of adopting a regional perspective that accounts for cross-jurisdictional travel patterns and the interconnected nature of transportation networks. By integrating high-fidelity mobility simulations, vehicle energy models, and environmental impact assessments, regional authorities can develop equitable, sustainable, and forward-looking strategies for facilitating the transition to electric mobility while prioritizing the well-being of disadvantaged communities.

\section{Building Spline Functions to Interpolate EV Demand Data}

Figures 1 and 2 illustrate the relationship between speed and two critical performance metrics for electric vehicles (EVs): range and energy consumption. This data is sourced from a Geotab study \cite{geotab} detailing how speed and temperature impact EV performance.

Figure 1 displays the driving range for sedans at a temperature of 68°F. 

\begin{figure}[h]
\centering
\includegraphics[width=0.45\textwidth]{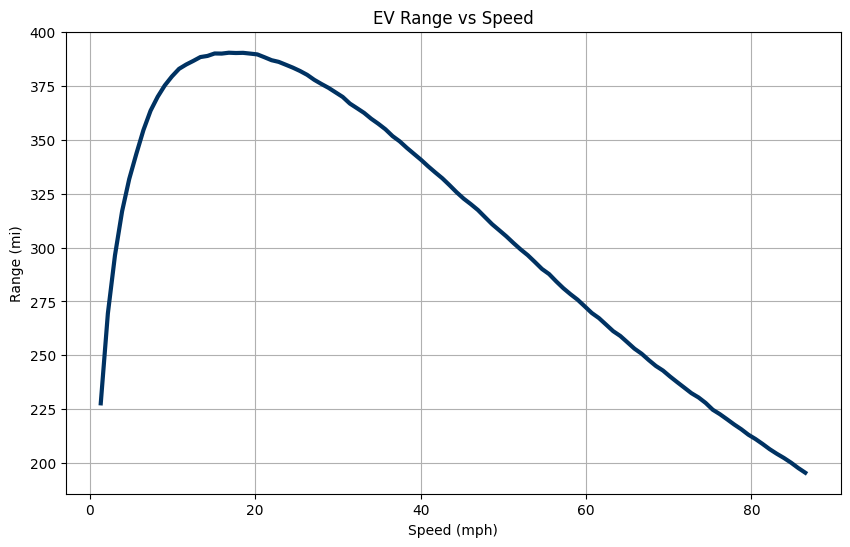}
\caption{EV Range vs. Speed}
\label{fig:range_speed}
\end{figure}

Figure 2 presents the model for energy consumption in Watt-hours per mile for sedans at a temperature of 68°F.

\begin{figure}[h]
\centering
\includegraphics[width=0.45\textwidth]{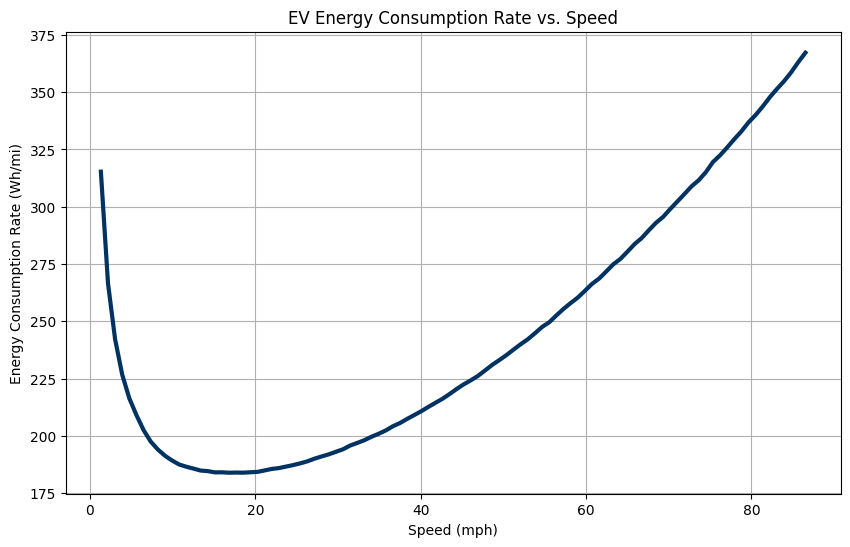}
\caption{EV Energy Consumption Rate vs. Speed}
\label{fig:energy_speed}
\end{figure}
A model for Figure 2 can be built as spline interpolation function for energy consumption estimates, represented as:
\begin{equation}
    E = f(v) = E_{rate}(v)
\end{equation}
\section{Mobiliti Simulation Data}
This section discusses the types of data generated by the Mobiliti simulation relevant to this project. Below are tables depicting various elements of the Mobiliti simulation data.

% LINKS
\begin{table}[ht]
\centering
\caption{Detailed Entry from Links}
\begin{tabular}{|c|c|c|c|}
\hline
\textbf{link\_id} & \textbf{node\_id\_in} & \textbf{node\_id\_out} & \textbf{...}\\ \hline
23592962 & 48501858 & 48501829 & ... \\ \hline
\end{tabular}
\\[5mm] % Space between tabulars
\begin{tabular}{|c|c|c|c|}
\hline
\textbf{...} & \textbf{free\_speed} & \textbf{length} & \textbf{capacity} \\ \hline
... & 11.111 m/s & 200.579 m & 1000 veh/h \\ \hline
\end{tabular}
\end{table}

% NODES
\begin{table}[h]
\centering
\caption{Sample entry from Nodes}
\begin{tabular}{|c|c|c|c|}
\hline
\textbf{node id} & \textbf{lon} & \textbf{lat} & \textbf{elev} \\
\hline
48501858 & -122.42562 & 37.61791 & 67.0 \\
\hline
\end{tabular}
\end{table}

% SPEEDS
\begin{table}[h]
\centering
\caption{Sample entry from Speeds}
\begin{tabular}{|c|c|c|c|c|c|}
\hline
\textbf{link id} & \textbf{00:00} & \textbf{00:15} & \textbf{...} & \textbf{23:30} & \textbf{23:45} \\
\hline
2840727 & 31.977 m/s & 31.977 m/s & ... & 29.743 m/s & 30.136 m/s \\
\hline
\end{tabular}
\end{table}

% LEGS
\begin{table}[h]
\centering
\caption{Sample entry from Legs}
\begin{tabular}{|c|c|c|c|c|}
\hline
\textbf{leg id} & \textbf{person id} & \textbf{orig node} & \textbf{dest node} & \textbf{...} \\
\hline
0 & 3 & 295 & 837 & ... \\
\hline
\end{tabular}
\\[5mm]
\begin{tabular}{|c|c|c|c|}
\hline
\textbf{...} & \textbf{start time} & \textbf{end time} & \textbf{duration} \\
\hline
... & 01:07:58.00 & 01:27:35.36 & 00:19:37.36 \\
\hline
\end{tabular}
\end{table}

% ROUTES
\begin{table}[h!]
\centering
\caption{Sample entry from Routes}
\begin{tabular}{|c|c|}
\hline
\textbf{leg id} & \textbf{route} \\
\hline
0 & [ 7000016319, 7000004316, 7000004317,  ... ] \\
\hline
\end{tabular}
\end{table}
\section{Energy Consumption Estimation}
Energy consumption estimation from the Mobiliti simulation utilizes two methods leveraging the interpolated functions for energy rate versus speed.

\subsection{Estimation Methods}
\subsubsection{Rough Order of Magnitude (ROM)}
This method uses total trip length and duration to determine the average speed of the trip. The ROM energy consumption is estimated by evaluating the energy consumption rate vs speed model with the average speed, as described below. This is a commonly used method.
\begin{equation}
    v_{{leg}_{avg}} = \frac{L_{{leg}_{total}}}{t_{{leg}_{total}}} 
\end{equation}
\begin{equation}
    E_{ROM} = L_{leg} \times E_{rate}\bigl(v_{{leg}_{avg}}\bigr)
\end{equation}
where
\begin{itemize}
    \item $L_{leg}$ is the total length of the leg
    \item $E_{rate}\bigl(v_{{leg}_{avg}}\bigr)$ is the energy consumption rate model
\end{itemize}

\par\vspace{\baselineskip} % Adds an extra line space
\subsubsection{Granular Bottom’s Up Estimate}
This method uses the route that each leg takes and computes the energy consumption along each link traversed within the route. The energy consumed for each link within the route depends on the time the person was on the link, and accounts for congestion at the individual link level, as described below.
\begin{equation}
    E_{total} = \sum_{i=1}^{n}E_i = \sum_{i=1}^{n}\biggl(L_i \times E_{rate}\bigl(v_{{leg}_{avg}}\bigr)\biggr)
\end{equation}
where,
\begin{itemize}
    \item $i$ is a link
    \item $E_i$ is the energy consumed on the link
    \item $L_i$ is the length of the link
    \item $E_{rate}\bigl(v_{{leg}_{avg}}\bigr)$ is the energy consumption rate model
\end{itemize}

\par\vspace{\baselineskip} % Adds an extra line space
\subsection{Comparison of Estimation Methods}
The top-down approach provides a broad, averaged estimate based on overall trip statistics. However, by incorporating variations in speed throughout the trip (as done in the granular approach), we observe an increase in the calculated energy usage, attributed to the detailed link-level data of the trip. The granular estimation method uses the average speed of all vehicles on each link over 15-minute intervals. Consequently, this method does not account for the specific speeds of individual vehicles.

\begin{figure}[h]
\centering
\includegraphics[width=0.45\textwidth]{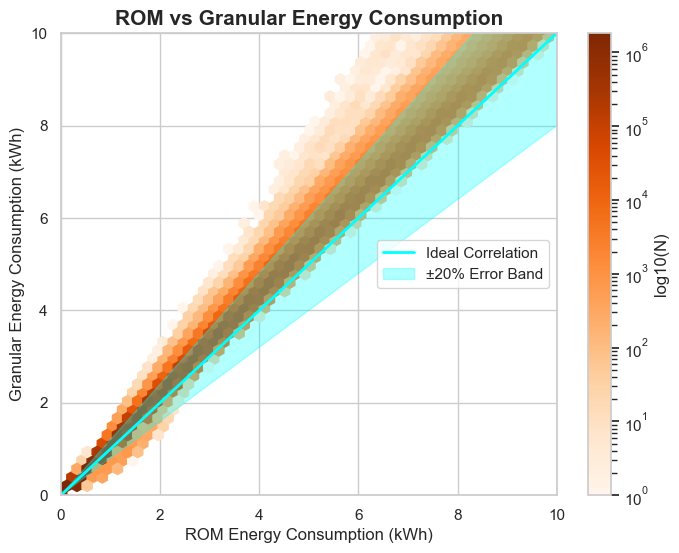}
\caption{ROM vs. Granular Energy Consumption Comparison}
\label{fig:rom_granular}
\end{figure}

The hexbin plot comparing ROM and Granular energy consumption estimates for over 19M trips in the Bay Area, shown in Figure~\ref{fig:rom_granular}  reveals several insights. The concentration of points along the diagonal in lower energy ranges suggests that the ROM method can be used for trips with lower energy demands. In contrast, for higher energy consumption scenarios, the Granular method provides a more detailed and possibly more accurate assessment, as it accounts for the specifics of each link along the route, including variable speeds.

From the hexbin plot, an important observation emerges for energy consumption levels exceeding 2 kWh. Notably, the +20\% error band around the ideal correlation line is closely aligned with the central concentration of the granular data relative to the ROM estimates. This alignment, combined with the absence of hexbin data points falling within the -20\% error band, strongly suggests that the granular method tends to account for up to 20\% more energy consumption compared to the ROM method. This finding implies that the granular method, which incorporates more detailed and specific data about individual link speeds over time, can capture additional energy expenditures that the ROM method, which uses broader, aggregated trip statistics, might overlook. This discrepancy highlights the enhanced sensitivity of the granular approach to factors that increase energy consumption, potentially making it a more accurate tool for scenarios where precision in energy estimation is critical.

Electric vehicles that estimate trip energy consumption using high-level statistics, such as total distance and average speed, may significantly underestimate actual energy needs. This underestimation can affect the predicted range of the vehicle, as such methods do not account for the specific route taken or variations in speed due to congestion on different segments of the route. This suggests a potential risk of range estimations being overly optimistic, especially in scenarios where traffic conditions cause considerable fluctuations in driving speeds as well as long trips.

\subsection{Dynamic RouteE (NREL)}
Building upon the foundational understanding that simple estimates based on high-level statistics might not accurately reflect real-world energy needs, the introduction of NREL's RouteE-powertrain model \cite{holden2020routee} into this analysis marks a significant enhancement. A new model, built as part of the DOE project "Big Data Solutions for Mobility Planning"\cite{bdsm}, incorporates critical factors previously overlooked, such as road grade and detailed powertrain dynamics, which impact fuel consumption. Preliminary results from comparing traditional Mobiliti energy models with RouteE estimates have demonstrated a closer alignment with actual fuel economies observed across different regions. This improvement is primarily attributed to RouteE's ability to account for the varying demands of road conditions and the inclusion of a broader spectrum of vehicle types, including hybrids and electric vehicles.

The RouteE model not only refines energy estimates but also integrates high-fidelity powertrain simulations and real-world driving data from connected vehicles navigating urban environments. This approach allows for a nuanced capture of the start/stop dynamics typical in congested city driving. By analyzing drive cycles over individual road links and incorporating outcomes from NREL's FASTSim software, RouteE provides a more granular view of energy impacts related to common urban driving events, such as stopping at traffic signals, idling due to congestion, and maneuvers, which are captured in the Mobiliti modeling platform.

Such comprehensive modeling, validated against actual on-road energy consumption data, ensures that the predictions are not just theoretical but practically viable, offering a more reliable basis for policy decisions and environmental impact assessments. 

\subsection{Comparison of Estimation Methods with Dynamic RouteE}
The integration of dynamic factors in energy consumption estimation significantly enhances model accuracy. This section presents two hexbin plots that visually compare the performance of RouteE against both the ROM and Granular methods.

\begin{figure}[h]
\centering
\includegraphics[width=0.45\textwidth]{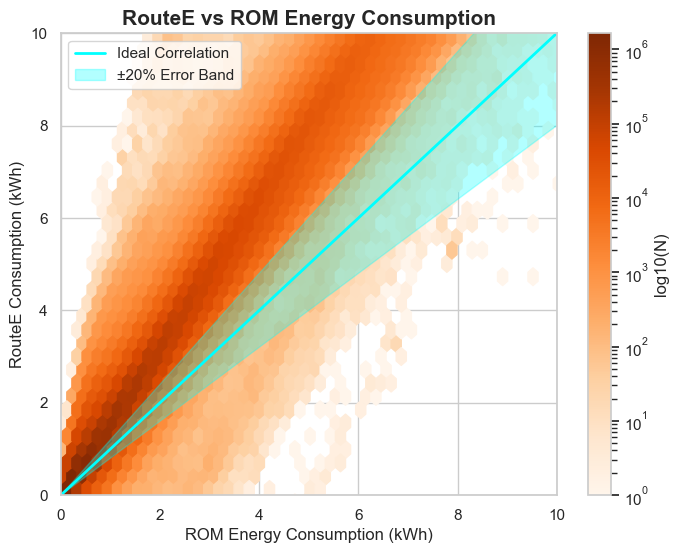}
\caption{Dynamic RouteE vs. ROM Energy Consumption Comparison}
\label{fig:routee_rom}
\end{figure}

The hexbin plots comparing RouteE to ROM and Granular methods reveal significant insights into the effectiveness of these energy consumption estimation models. Figure~\ref{fig:routee_rom} displays a considerable spread of data points, which indicates a variable level of agreement between the two models. This variability is likely due to the ROM method’s simpler assumptions, which do not account for the more dynamic factors included in RouteE. 

\begin{figure}[h]
\centering
\includegraphics[width=0.45\textwidth]{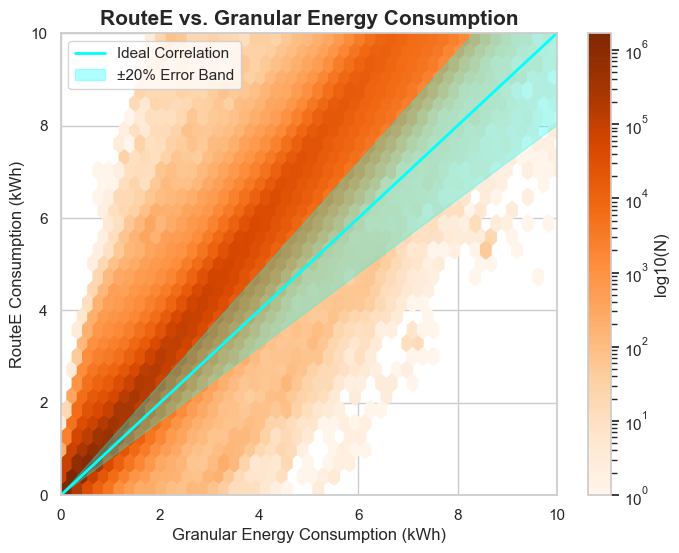}
\caption{Dynamic RouteE vs. Granular Energy Consumption Comparison}
\label{fig:routee_granular}
\end{figure}

Compared to the ROM, Figure~\ref{fig:routee_granular} shows a tighter clustering of data points around the ideal correlation line. This improved correlation across the energy consumption spectrum suggests that incorporating speeds at individual link levels offers a more accurate reflection of energy dynamics.  The slightly better correlation observed in Figure~\ref{fig:routee_granular} compared to Figure~\ref{fig:routee_rom} suggests that modeling energy consumption at the granularity of individual road links (as done in the Granular method) is more effective. This approach moves closer to replicating the sophisticated dynamics captured by RouteE, particularly in urban settings where driving conditions can vary dramatically over short distances. 

These findings suggest that for accurate energy consumption estimation, particularly in complex urban environments, models need to incorporate as many real-world dynamics as possible. Mobiliti generates the routes for all of the 19M+ trips in the Bay Area simulation, and consequently can build better estimates for energy consumption across road networks leveraging different models like Dynamic RouteE or other estimation methods. Obviously, the application of the model will dictate the level of complexity of these models, e.g. Dynamic RouteE being dependent on large machine learning models, that is necessary. Currently, Mobiliti uses the Granular energy consumption model which lies between the ROM model and complex RouteE model in its ability to predict consumption. Further investigation could show the sensitivities as a function of the use case.

\section{Demand Density}
Designing infrastructure for battery electric vehicles (BEVs) necessitates synchronizing two perspectives: the electric grid's capacity to service energy demand at locations where drivers require charging, and understanding the geospatial locations where vehicles will need to recharge. This dual optimization problem is highly complex and may not reflect real world scenarios. The present analysis considers only the dynamics of vehicle behavior and predicts energy consumption (using the Granular model) to forecast where BEVs will likely require charging. We evaluate five battery level thresholds (10 kWh, 20 kWh, 30 kWh, 40 kWh, and 50 kWh) to account for the variety of vehicle types currently available on the market. By modeling vehicle energy consumption patterns, we aim to inform the optimal siting of charging infrastructure to meet the anticipated spatial demand from BEVs across different battery capacities.

To strategically evaluate the demand for electric vehicle (EV) charging facilities across the Bay Area, we employed a detailed energy consumption model using the granular approach. This model calculates and tracks energy usage for over 8 million unique individuals participating in the simulation, each characterized by multiple activity-based trip segments, or legs, accounting for over 19 million trips in total. 

Each participant's journey is analyzed leg-by-leg, taking into account the specific route and prevailing traffic conditions. For every link in a trip, the model computes the energy consumed based on the average speed during a 15-minute window and the length of the link, using the energy consumption rate (Wh/mi) versus speed function. The energy consumption for each link is then aggregated across all legs of a person's itinerary.

Throughout their journeys, individuals' cumulative energy consumption is continuously monitored against predefined thresholds set at 10 kWh increments, ranging from 10 kWh to 50 kWh. These thresholds are reflective of the varying battery capacities found in different types of electric vehicles. When a person's cumulative energy use surpasses one of these thresholds, the location and the corresponding energy level are recorded.

For each of the over 19 million trips in our model, we identified the locations where vehicles reach the five battery level thresholds (10 kWh, 20 kWh, 30 kWh, 40 kWh, and 50 kWh). Shorter trips failing to meet a given threshold were filtered from that dataset. Longer trips surpassing each threshold were retained as potential candidates for recharging at those locations. 

The culmination of this analysis is presented through a series of demand density plots, each corresponding to a specific energy threshold. These plots utilize 3D hexbins, each 1 km in size, where the height of a hexbin is proportional to the number of times a threshold is exceeded within that area. This method provides a clear, visual representation of potential hotspots for EV charging demand across the region.

Figures 1-5 depict the geospatial distribution of locations where vehicles attain the respective threshold levels. A diminishing number of locations are observed as the thresholds increase, reflecting the higher energy requirements for longer-range trips. The 50 kWh threshold represents trips necessitating the highest battery capacities in our analysis. This multi-threshold approach enables modeling the spatial demand for charging infrastructure across various vehicle ranges and battery capacities.

\begin{figure}[h]
\centering
\includegraphics[width=0.45\textwidth]{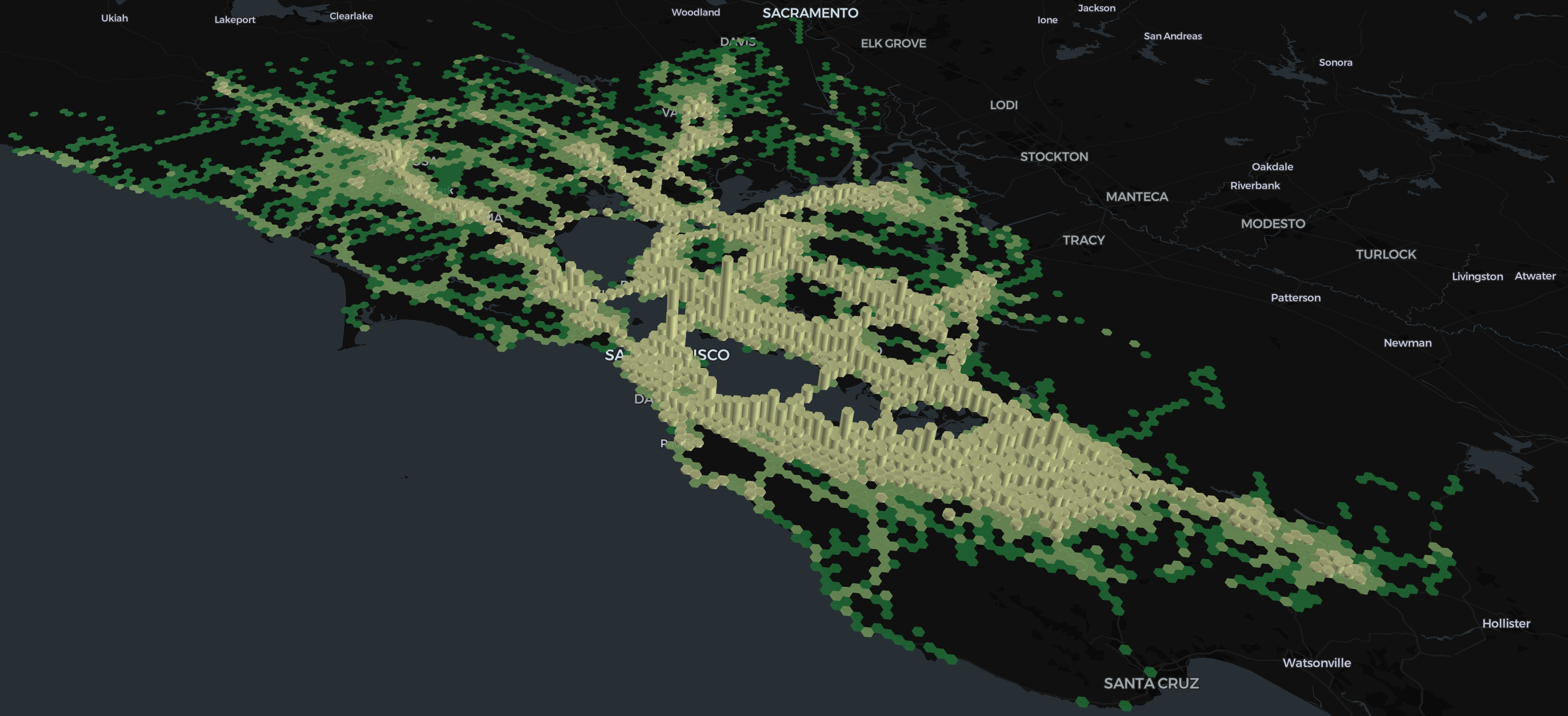}
\caption{10 kwh Threshold Locations}
\label{fig:10kwh_density}
\end{figure}

\begin{figure}[h]
\centering
\includegraphics[width=0.45\textwidth]{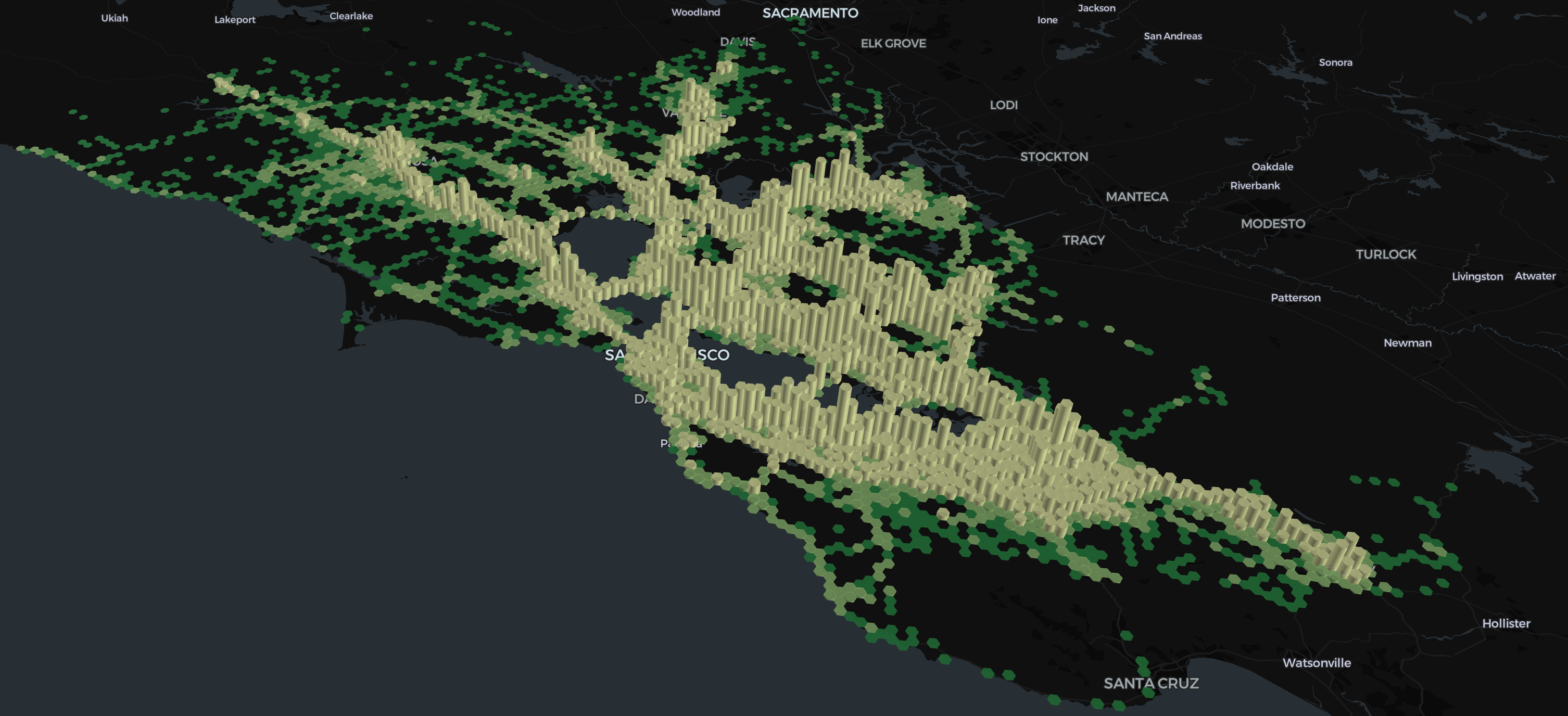}
\caption{20 kwh Threshold Locations}
\label{fig:20kwh_denisty}
\end{figure}

\begin{figure}[h]
\centering
\includegraphics[width=0.45\textwidth]{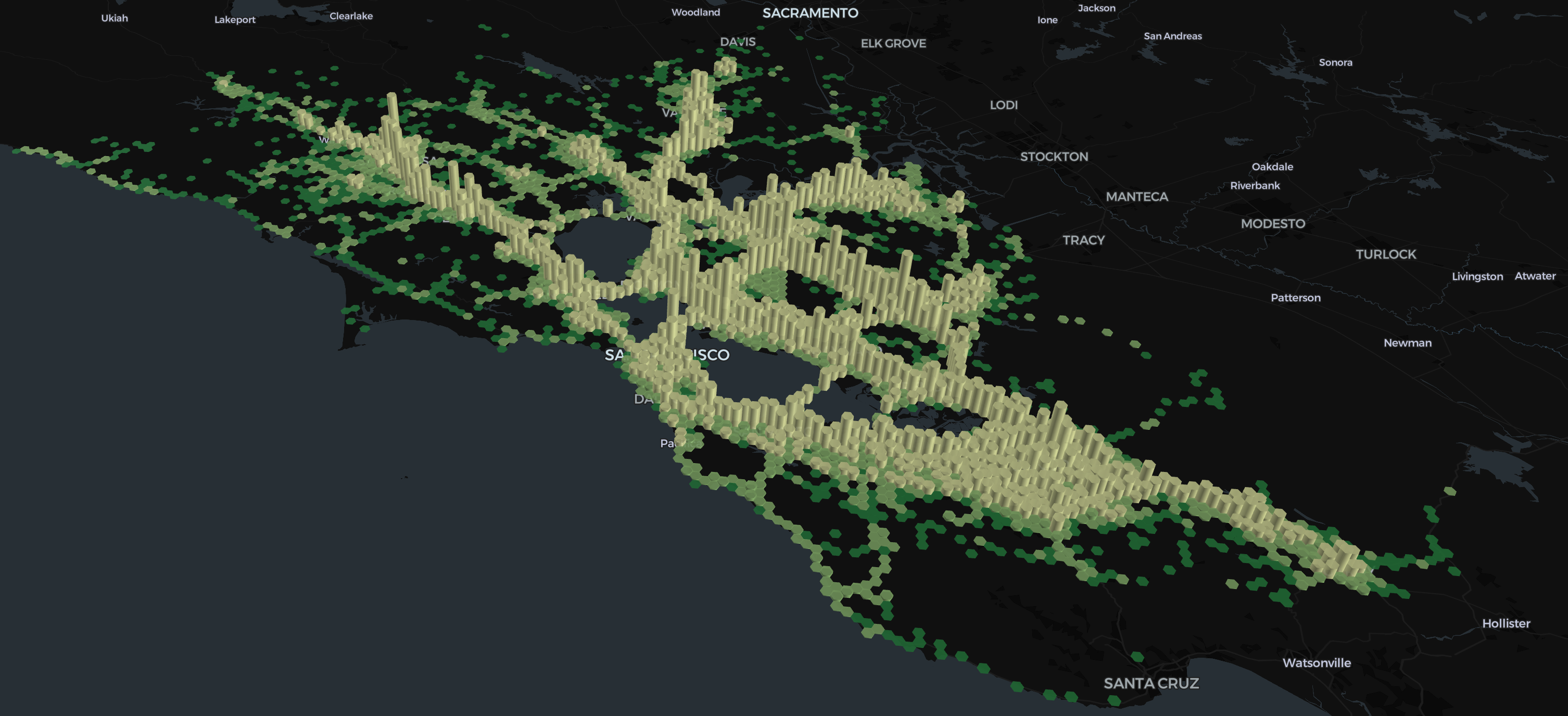}
\caption{30 kwh Threshold Locations}
\label{fig:30kwh_denisty}
\end{figure}

\begin{figure}[h]
\centering
\includegraphics[width=0.45\textwidth]{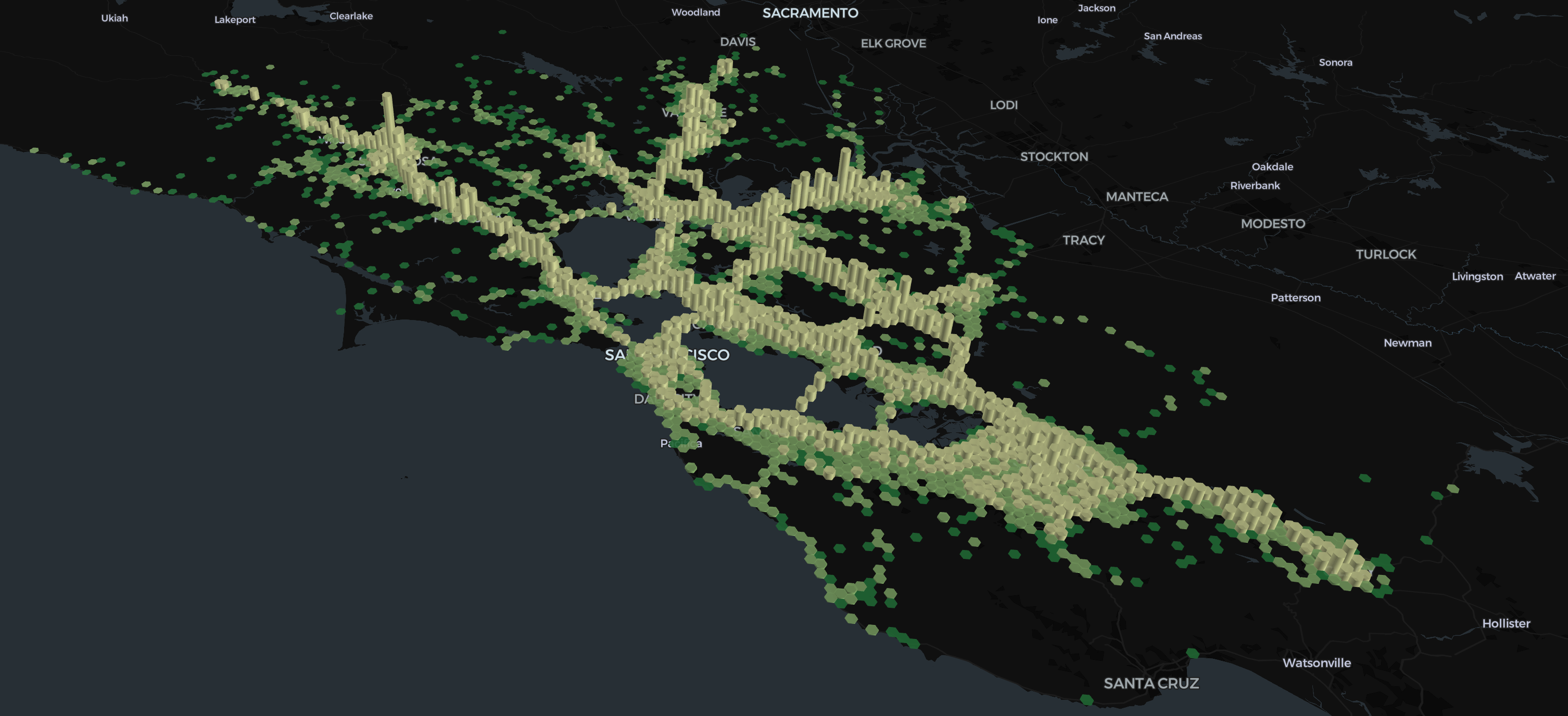}
\caption{40 kwh Threshold Locations}
\label{fig:40kwh_denisty}
\end{figure}

\begin{figure}[h]
\centering
\includegraphics[width=0.45\textwidth]{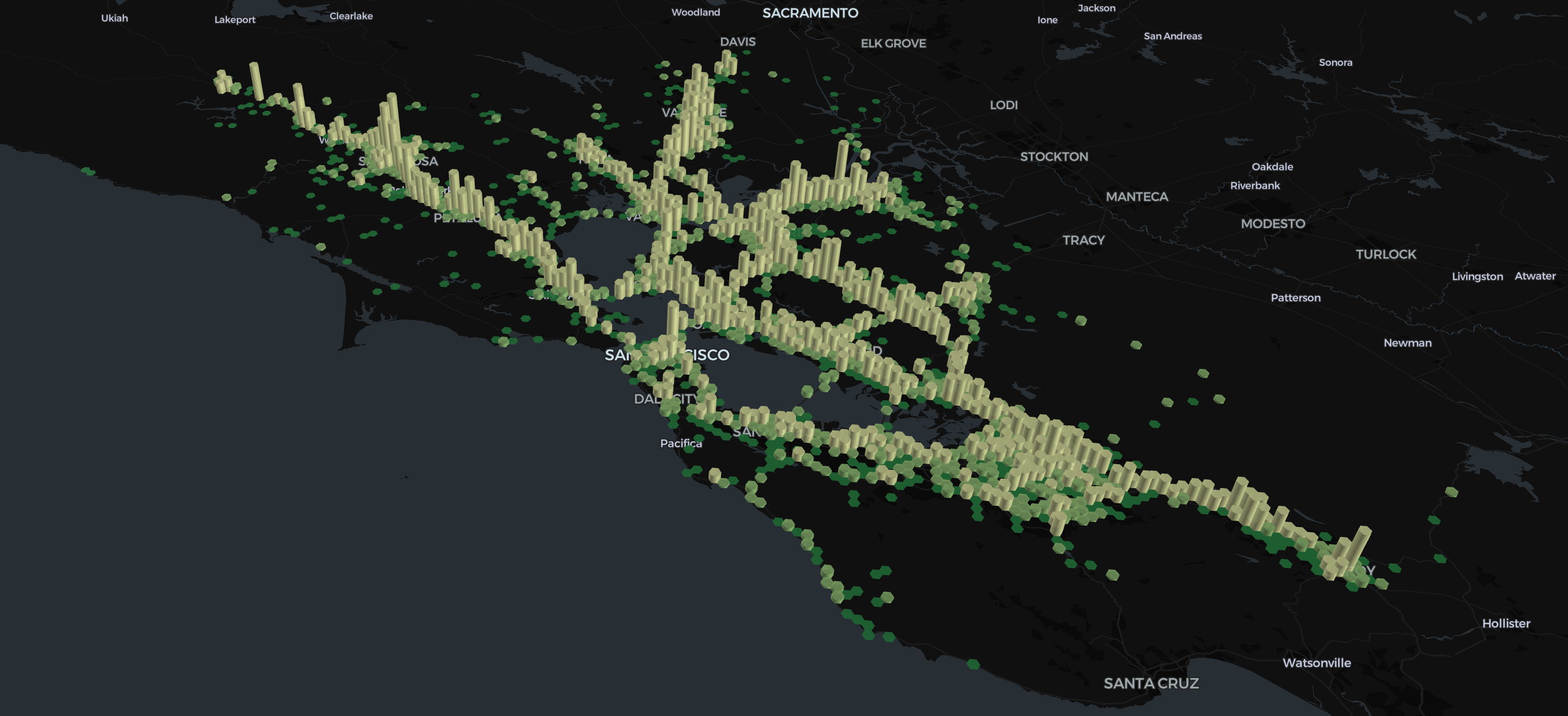}
\caption{50 kwh Threshold Locations}
\label{fig:50kwh_denisty}
\end{figure}
\section{Impacts of BEV Adoption on Equity Priority Communities}
\subsection{Context and Health Impacts}
Near-highway communities experience disproportionate exposure to traffic emissions, leading to adverse health effects including respiratory issues (asthma, bronchitis, COPD), cardiovascular diseases (heart attacks, strokes), cancer risk (lung, leukemia), neurological impacts (cognitive decline, neurodevelopmental disorders), and reduced birth weights/developmental delays. Key pollutants include NOx, PM, VOCs, benzene, formaldehyde, PM2.5, NO2, and PAHs. Low-income and minority communities are disproportionately impacted due to historical residential patterns and limited healthcare access. 

\subsection{Mitigation and Analysis Focus}
Mitigation strategies involve reducing emissions, improving transportation options, enhancing green buffers, prioritizing environmental justice in urban planning, and community empowerment. For this analysis, we focus solely on the reduced emission benefit associated with EV adoption.

\subsection{EV Demand Density Considerations}
\begin{figure}[h]
\centering
\includegraphics[width=0.45\textwidth]{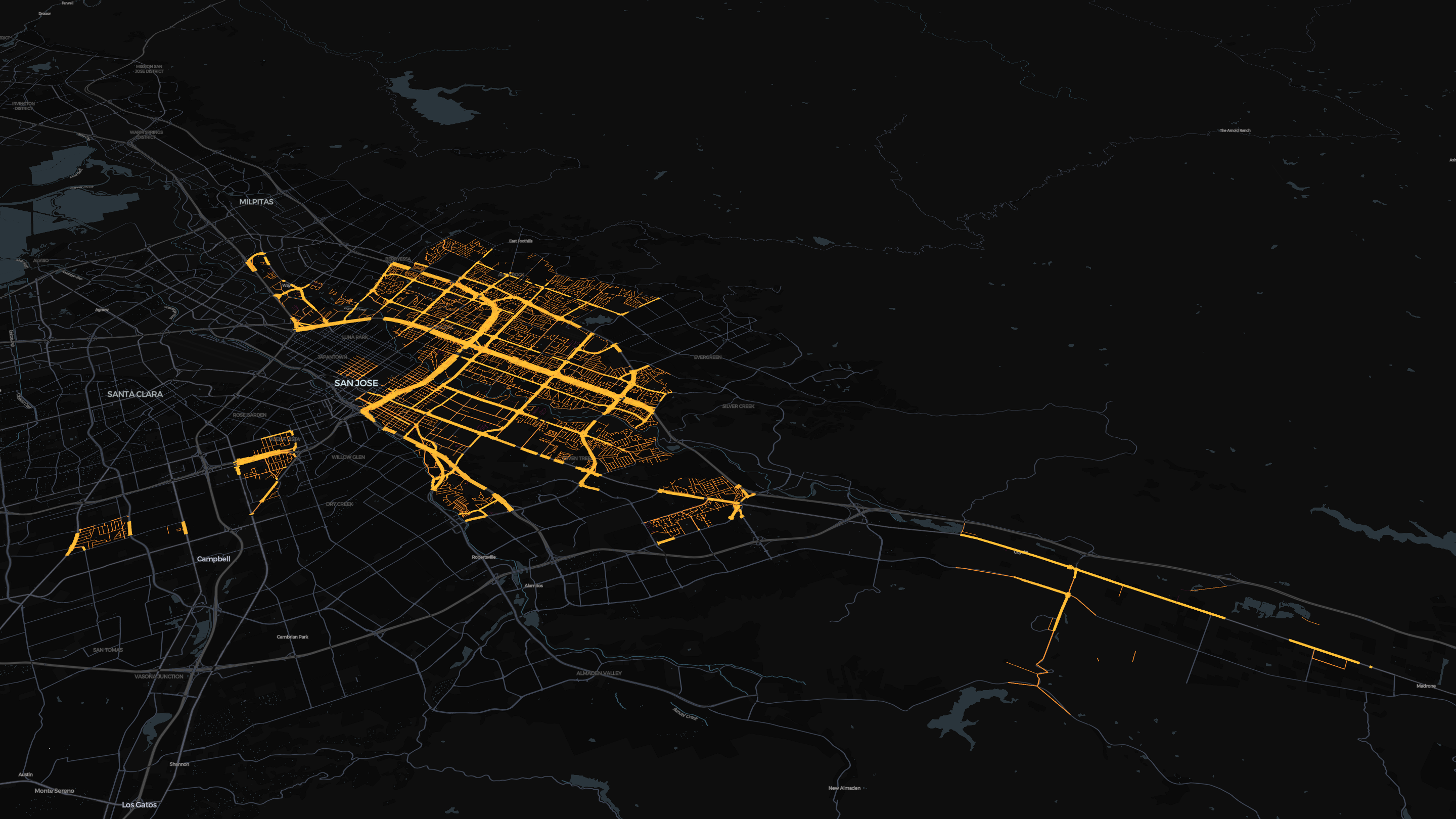}
\caption{Equity Priority Communities' Links in the San Jose Area}
\label{fig:epc_links}
\end{figure}

In the demand density analysis, the original 19M+ trips associated with the Bay Area travel profile, are filtered to remove localized mobility. Trips not meeting a 10kWh minimum threshold were filtered out to exclude short trips readily completed by EVs. Consequently, when analyzing locations of EVs exceeding this threshold, we observe EV trips involving significant distances, likely originating from external communities and passing through our Equity Priority Communities (EPCs) area of interest (Figure~\ref{fig:epc_links}). 

\begin{figure}[h]
\centering
\includegraphics[width=0.45\textwidth]{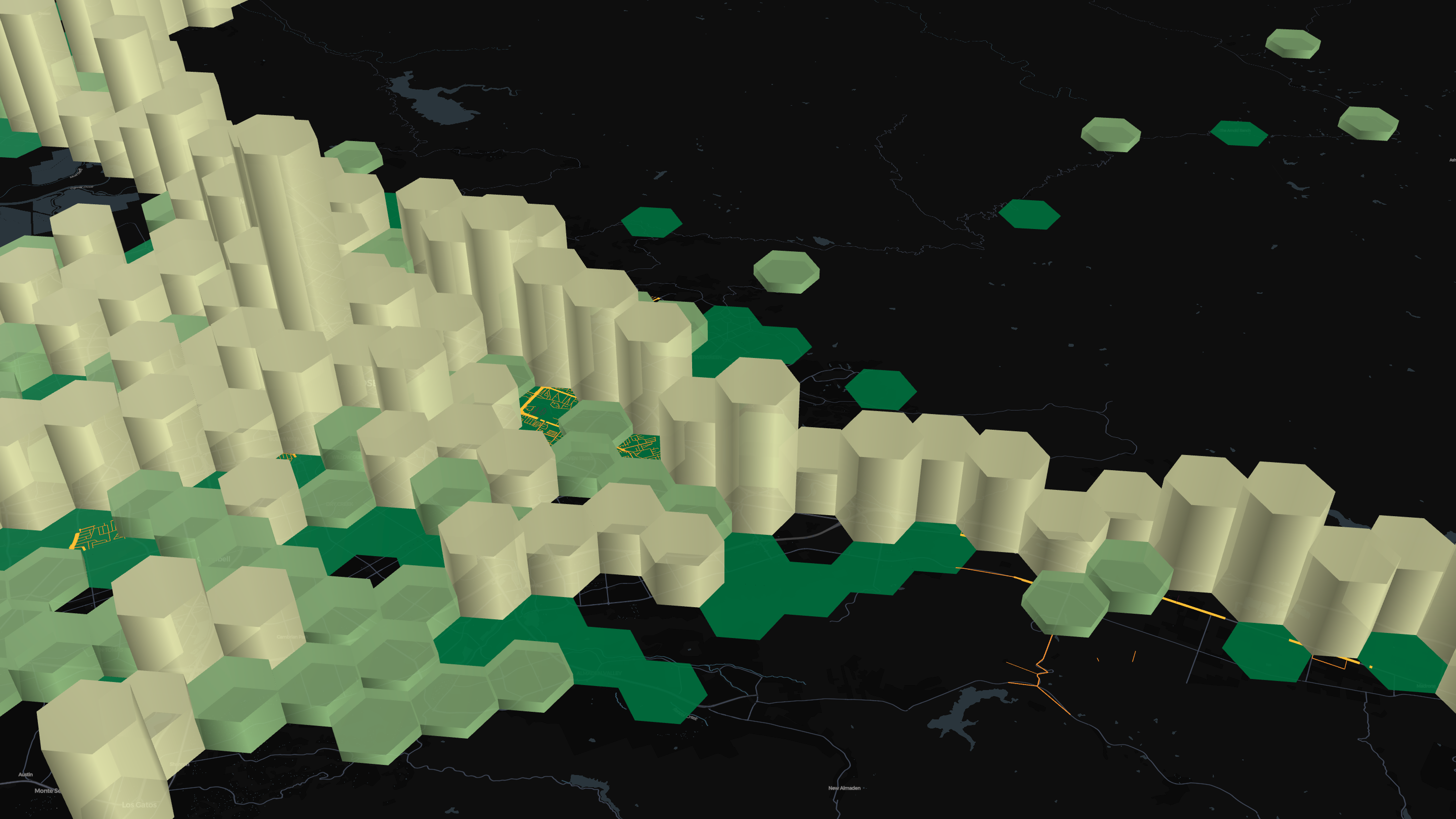}
\caption{50 kWh Threshold Locations (3D) in and around EPC Links in San Jose}
\label{fig:epc_links_50kwh_3d}
\end{figure}

As an example, Figure~\ref{fig:epc_links_50kwh_3d} shows an overlay of the 50 kWh threshold hexbins in 3D with the EPC links in the San Jose Area.  There is a significant presence of 50 kWh thresholds in proximity to the EPC links. This pattern indicates that many trips passing through or near these links originate from distant locations.

While disregarding safety concerns associated with pass-through traffic, these EV trips, along with those on adjacent freeways, can contribute to significant reductions in NOx and CO2 levels for Equity Priority Communities (EPCs). By considering only longer-range EV trips, we capture potential emission reductions from BEVs traveling through EPCs, rather than local trips within the community itself.

% \begin{figure}[h]
% \centering
% \includegraphics[width=0.45\textwidth]{Figures/Fig13_EPC-links-10kWh-2D.png}
% \caption{10 kWh Threshold Locations (2D) in and around EPC Links in San Jose}
% \label{fig:epc_links_10kwh_2d}
% \end{figure}

% \begin{figure}[h]
% \centering
% \includegraphics[width=0.45\textwidth]{Figures/Fig14_EPC-links-10kWh-3D.png}
% \caption{10 kWh Threshold Locations (3D) in and around EPC Links in San Jose}
% \label{fig:epc_links_10kwh_3d}
% \end{figure}

% \begin{figure}[h]
% \centering
% \includegraphics[width=0.45\textwidth]{Figures/Fig15_EPC-links-20kWh-3D.png}
% \caption{20 kWh Threshold Locations (3D) in and around EPC Links in San Jose}
% \label{fig:epc_links_20kwh_3d}
% \end{figure}

% \begin{figure}[h]
% \centering
% \includegraphics[width=0.45\textwidth]{Figures/Fig16_EPC-links-30kWh-3D.png}
% \caption{30 kWh Threshold Locations (3D) in and around EPC Links in San Jose}
% \label{fig:epc_links_30kwh_3d}
% \end{figure}

% \begin{figure}[h]
% \centering
% \includegraphics[width=0.45\textwidth]{Figures/Fig17_EPC-links-40kWh-3D.png}
% \caption{40 kWh Threshold Locations (3D) in and around EPC Links in San Jose}
% \label{fig:epc_links_40kwh_3d}
% \end{figure}

\subsection{Methodology for Estimating Fuel and Emissions Reduction in EPCs}
In our study, we utilized a Monte Carlo simulation to model the potential impact of BEV adoption on localized emissions. We assigned EVs to 37\% of eligible persons, reflecting the current market penetration rates \cite{ThirtySevenPercentEV}. This assignment excluded individuals whose total trip distance in terms of the sum of all of their legs exceeded 300 miles. Our focal area for this analysis was a set of links defining a Equity Priority Communities (EPCs) in the San Jose area.

The Monte Carlo approach repeatedly simulated (1,000 iterations) the assignment of EVs to eligible individuals and analyzed trips that passed through the EPC's designated links. For each EV trip that passed through the EPC, we calculated the reduction in fuel consumption using the RouteE model, which integrates with our Mobiliti simulator to estimate fuel consumption in liters of gasoline.

\subsection{Impact Quantification}
\begin{figure}[h]
\centering
\includegraphics[width=0.45\textwidth]{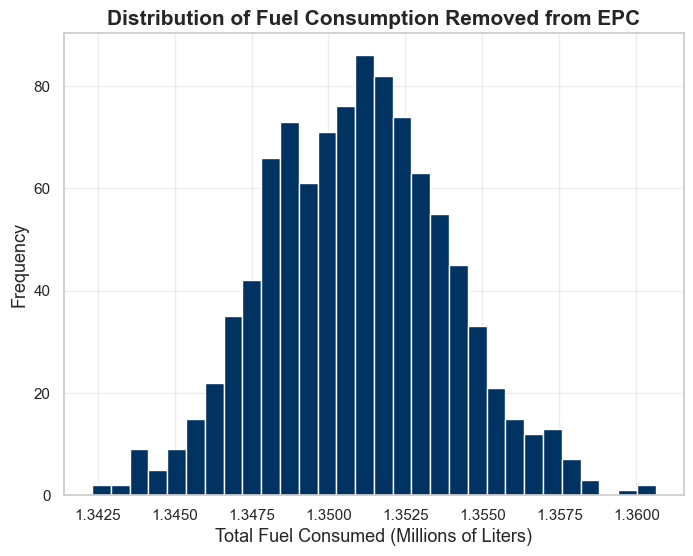}
\caption{Distribution of fuel consumption removed in and around Equity Priority Communities in San Jose from Monte Carlo simulation of EV penetration in the Bay Area}
\label{fig:epc_fuelremoved}
\end{figure}

The cumulative impact of these trips on local air quality was quantified by summing the fuel savings for all relevant EV trips. This data was then visualized in a histogram, Figure~\ref{fig:epc_fuelremoved}, depicting the distribution of total fuel removed from the EPC due to BEV adoption. The average fuel consumption removed is approximately 1.35M liters.

In establishing the initial National Program fuel economy standards for model years 2012-2016, the EPA and the Department of Transportation, on May 7, 2010, adopted a unified conversion factor. This factor, documented in the Federal Register of 2010, sets the CO2 emissions from gasoline combustion at 8,887 grams per gallon. This conversion is based on the complete conversion of carbon in gasoline to CO2, a principle also supported by IPCC guidelines from 2006. The conversion calculation used is straightforward: 8,887 grams of CO2 per gallon translates to approximately 0.008887 metric tons of CO2 per gallon.

\begin{figure}[h]
\centering
\includegraphics[width=0.45\textwidth]{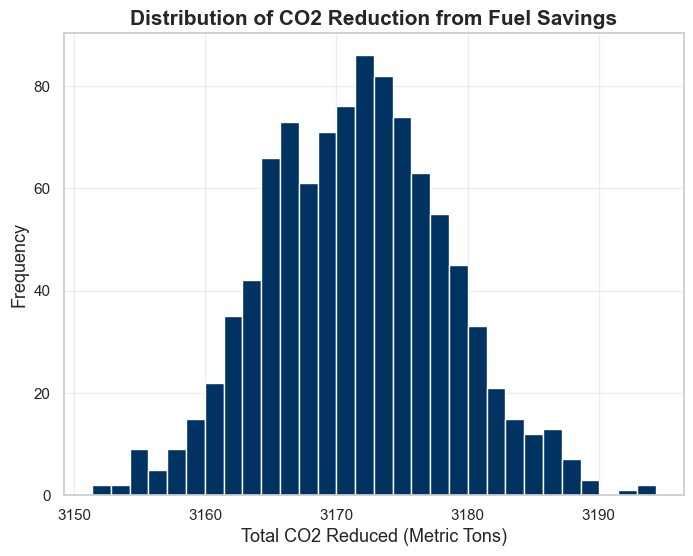}
\caption{Distribution of CO2 removed in and around Equity Priority Communities in San Jose from Monte Carlo simulation of EV penetration in the Bay Area}
\label{fig:epc_CO2removed}
\end{figure}

Utilizing the outcomes from our Monte Carlo simulation, the estimated fuel savings was converted into CO2 reductions, expressed in metric tons. This was achieved by applying the established conversion factor of 0.008887 metric tons of CO2 per gallon of gasoline saved. The result is captured in Figure~\ref{fig:epc_CO2removed}, which shows the distribution of CO2 reduction in an around the EPC area of interest. This methodology allows us to quantify the environmental benefit of transitioning to BEVs in terms of CO2 emission reduction. The average reduction is estimated to be 3,173 metric tons of CO2.

\subsection{Conclusion}
This detailed analysis provides valuable insights into how targeted BEV adoption strategies could substantially reduce harmful emissions in Equity Priority Communities, thereby alleviating some of the disproportionate environmental burdens these communities face. The integration of high-fidelity simulations with real-world data ensures that our findings are both scientifically robust and practically relevant, offering a reliable basis for future policy and planning decisions aimed at enhancing community health and environmental quality.

\section{CONCLUSIONS}

The shift towards widespread adoption of battery electric vehicles (BEVs) and the development of necessary charging infrastructure requires a regional approach transcending jurisdictional boundaries. This study highlights the vital role of high-speed, large-scale regional transportation simulation models in iteratively optimizing charging station networks. Using the Mobiliti platform's parallel discrete event simulation capabilities, we efficiently analyzed over 19 million trips in the San Francisco Bay Area. This framework enabled the evaluation of multiple scenarios, refining vehicle demand densities and assessing impacts on emissions and environmental justice communities. As municipalities and regional authorities engage in infrastructure planning, access to computable simulation models is crucial. These models will allow exploration of siting options, assessment of community impacts, and adaptation to emerging mobility patterns and technology advancements, ensuring investments are future-proofed and maximize long-term benefits.

Equally important is the emphasis on equity throughout the planning process. By quantifying emission reductions in equity priority communities resulting from BEV adoption, this study demonstrates the potential for strategic charging infrastructure deployment to address environmental justice issues. 

Advancements in computing, data integration, and simulation methodologies will further support equity-driven design of charging networks. Collaboration among transportation authorities, urban planners, community stakeholders, and technology providers is essential to leverage these tools, ensuring the transition to electric mobility occurs equitably, sustainably, and responsibly.

\addtolength{\textheight}{-12cm}   % This command serves to balance the column lengths

%%%%%%%%%%%%%%%%%%%%%%%%%%%%%%%%%%%%%%%%%%%%%%%%%%%%%%%%%%%%%%%%%%%%%%%%%%%%%%%%

\bibliography{references.bib}
\bibliographystyle{IEEEtran}

%%%%%%%%%%%%%%%%%%%%%%%%%%%%%%%%%%%%%%%%%%%%%%%%%%%%%%%%%%%%%%%%%%%%%%%%%%%%%%%%

%%%%%%%%%%%%%%%%%%%%%%%%%%%%%%%%%%%%%%%%%%%%%%%%%%%%%%%%%%%%%%%%%%%%%%%%%%%%%%%%

\section*{ACKNOWLEDGMENT}
The Mobiliti Regional Scale Transportation modeling platform is the results of many contributors, particularly Cy Chan of Lawrence Berkeley National Laboratory. Mobiliti development was sponsored by the U.S. Department of Energy (DOE) Vehicle Technologies Office (VTO) under the Big Data Solutions for Mobility Program, an initiative of the Energy Efficient Mobility Systems (EEMS) Program. The following DOE Office of Energy Efficiency and Renewable Energy (EERE) managers played important roles in establishing the project concept, advancing implementation, and providing ongoing guidance: David Anderson, Prasad Gupte and Avi Mersky.
%%%%%%%%%%%%%%%%%%%%%%%%%%%%%%%%%%%%%%%%%%%%%%%%%%%%%%%%%%%%%%%%%%%%%%%%%%%%%%%%
\end{document}